# An intuitive scheme for the restoration of entanglement for polarization-entangled photons emitted from quantum dots with non-vanishing fine structure splitting

Simone Varo[1*], Gediminas Juska[1] and Emanuele Pelucchi[1]

1    Tyndall National Institute, University College Cork, Dyke Parade, Cork, Republic of Ireland
*Simone.varo@tyndall.ie

**Abstract:** Generation of polarization-entangled photons from quantum dots via the biexciton-exciton recombination cascade is complicated by the presence of an energy spitting between the intermediate excitonic levels, which severely degrades the quality of the entangled photon source. In this paper we present a novel, conceptually simple and straightforward proposal for restoring the entanglement of said source by applying a cascade of time-dependent operations on the emitted photons. This is in striking contrast with the techniques usually employed, that act on the quantum emitter itself in order to remove the fine structure splitting at its root. The feasibility of the implementation with current technology is discussed, and the robustness of the proposed compensation scheme with respect to imperfections of the experimental apparatus is evaluated via a series of Monte Carlo simulations.

**Introduction**

Quantum dots (QDs) have emerged as one of the most promising candidates for on-demand generation of non-classical light, due to their ability to generate single photons[1–3] and even entangled ones[4–7], especially via the biexciton-exciton recombination cascade mechanism[8].

The presence of a fine structure splitting (FSS) between the intermediate excitonic levels[9,10] has been for years the most severe obstacle in the generation of high quality polarization-entangled photon pairs: in a textbook example of Noether's theorem, the degeneracy of the levels is promptly lifted not only due to shape anisotropy but even because of disorder in the alloy of the barriers of the dot, or strain and piezoelectrically induced electric fields which reduce the symmetry of the confinement potential[11]. This causes a residual FSS to appear even in highly symmetrical dots[12], in spite of significant improvements in growth techniques that helped reducing it significantly in recent years[13–17].

Several approaches have been proposed to tackle this issue at its root[18], i.e. relying on the application of external tuning knobs in the form of the electric[19] and optical Stark effect[20], and magnetic[21] or strain[22–24] fields in order to restore the symmetry and remove the source of the problem, or by annealing of the quantum dot to mitigate the severity of the effect[25]. While several attempts have been made to integrate the most successful of these solutions on semiconductor platforms in order to achieve compact and scalable entangled photon sources[24], this usually implies a significant increase in fabrication complexity, a limited yield in terms of working dots and devices, or brings in new problems such as the quenching of the photoluminescence when electric field tuning is employed.

Most of these attempts undertaken to restore the quality of the emitted entangled photons, however, start from the assumption that the degradation of the entanglement properties of the source is an irreversible process, and thus focus on removing the FSS. A posteriori solutions to mitigate FSS effects after the photons have been emitted are possible, but only a few proposal in this sense have emerged so far[26–31] and, to our knowledge, none of them has been experimentally implemented at the time of this paper being written. Moreover, it looks like most of these proposals start by identifying the energy difference of the photons, inherited by the excitonic FSS of the quantum dot, as the issue to tackle: the underlying assumption is that such a difference makes the two possible decay channels of the biexciton-exciton cascade distinguishable[6,28,32,33], thus breaking quantum superpositions in the final state. Indeed, the idea that the existence of a "which-path" information is the root of polarization entanglement degradation was widely accepted in the community in the past, and it would stand to reason in such a framework that performing a quantum erasure by restoring energy degeneracy of the photons before any measurement is done would yield perfectly entangled states. However, the now widely accepted mechanism via which FSS impacts on the generation of entangled photons via the biexciton-exciton recombination cascade was later discussed in detail by Stevenson et al. in their seminal paper[34], and will be briefly recalled in order to establish the theoretical framework and notation used in the rest of our paper.

Let |XX> indicate the biexcitonic (electronic) state of the quantum dot, and $|X_H>$ and $|X_V>$ the two intermediate excitonic states, with energies $E_X^H$ and $E_X^V$, as shown in figure 1. If the emitter is populated with a biexciton (e.g. via a two-photon resonant pumping scheme), it will decay after a time $t_1$ leading to an exciton-photon entangled state of the form $|\psi> \alpha(|H_{XX}X_H>+|V_{XX}X_V>)$, where $|H_{XX}>$ and $V_{XX}>$ respectively denote a horizontally and vertically polarized photon emitted during the biexciton to exciton transition.

Before the completion of the recombination cascade, marked by the exciton to ground state transition after a temporal delay $t_2$, a different energy of the two intermediate levels, that is, the presence of an FSS, will lead to a different time evolution of the excitonic states as per the time-dependent solution of the Schrödinger equation:

$$|\Psi(t)>= \alpha(|H_{XX}X_H> * e^{i(E_X^H*t)/\hbar} + |V_{XX}X_V> * e^{i(E_X^V*t)/\hbar}) \qquad (1)$$

With the resulting phase accumulated in this interval finally transferred to the two photons state resulting from the overall cascade, which, neglecting global phase terms can be written as:

$$|\psi> = \alpha (|H_{XX}H_X> + e^{i(FSS*t_2)/\hbar}|V_{XX}V_X>) \qquad (2)$$

With $|H_X>$ and $|V_X>$ being the horizontally and the vertically polarized photons emitted during this second transition.

As it can be seen, the state is still an entangled one, albeit not the Bell state that would be produced should the FSS be zero. The random nature of the interval $t_2$ that precedes the recombination of the exciton implies that while the dot would always emit entangled photon pairs, the state each pair would be entangled to will be a (different) random one, effectively mimicking, in a statistical sense, classical light, and making such a source relatively useless for practical applications should the FSS be sufficiently large to broaden the distribution of the generated states[35]. Post selection techniques based on time-gating are possible, and allow to discard of photon pairs generated for longer values of the recombination time $t_2$ where the phase build-up causes a significant drift of the state from the maximally entangled Bell's one: the obvious downside is that a trade-off has to be reached between the desired level of entanglement and the amount of acceptable losses. For instance, sometimes the fraction of acceptable photons drops to a few percentage points[36] of the overall emission from the quantum dot.

However, as each emitted state is already an entangled one, the problem of restoring the quality of the entanglement of the source could be reformulated as the search for a unitary transformation that merely transforms, a posteriori, any of these states to a predetermined one, thus removing statistical fluctuations.

As discussed before, most approaches present in the literature have focussed on shifting the energy of the two polarization states of the emitted photons so that they become degenerate after the emission. The approach we hereby propose however, builds directly on the theoretical framework developed by Stevenson and that we have recalled, and starts by considering the difference in phase and not in energy between the two excitonic photon states as the issue to tackle.

**Results and discussion**

### Determination of the conditions for entanglement restoral

Let's assume that both the biexciton and the exciton photons fly across a device which has the ability to impart a phase which is both different depending on the polarization and wavelength and changing linearly in time, so that we can write the phases introduced on either polarization of the photons as:

$$\Phi_{V_{XX}}(t) = K_{V_{XX}} * t + \Phi^0_{V_{XX}} \qquad \Phi_{V_X}(t) = K_{V_X} * t + \Phi^0_{V_X}$$

$$\Phi_{H_{XX}}(t) = K_{H_{XX}} * t + \Phi^0_{H_{XX}} \qquad \Phi_{H_X}(t) = K_{H_X} * t + \Phi^0_{H_X}$$

(3)

With the various K representing the slopes of the introduced phases for a given polarization and wavelength of a photon and the $\Phi^0$ terms indicating the constant (that is, time independent) phases introduced by the compensation system.

In general, if we assume the zero of our time scale to correspond with the initialization of the biexciton, the state resulting from such an operation can be written as a function of the time intervals $t_1$ and $t_2$ (with $t_1$ and $t_2$ being the random times before the emission of the biexciton and of the exciton photons, respectively) as:

$$|H_{XX}> e^{(i*\Phi_{H_{XX}}*(t_1+t_{prop}^{XX}-t_{start}^{XX}))} \otimes |H_X> e^{(i*\Phi_{H_X}(t_1+t_2+t_{prop}^{X}-t_{start}^{X}))} +$$
$$+|V_{XX}> e^{(i*\Phi_{V_{XX}}(t_1+t_{prop}^{XX}-t_{start}^{XX}))} \otimes |V_X> e^{(i*\Phi_{V_X}(t_1+t_2+t_{prop}^{X}-t_{start}^{X}))} * e^{(i*FSS*t_2/\hbar)}$$

(4)

Where $t_{prop}^{X,XX}$ (*prop* being short for propagation) is the time of flight of the biexciton (exciton) photon from the quantum dot to the entrance of the compensation system, and $t_{start}^{X,XX}$ is the time at which the ramping of the differential phases begin. With our choice of the zero of the time scale, we will thus have that if $t_{prop}^{XX} = t_{start}^{XX}$, the phase ramp for the biexciton photon will start when an XX photon emitted at t=0 (immediately after the dot has been excited) will reach the compensation system.

Substituting relations (3) and collecting all terms we can get an overall state of the form:

$$|H_{XX}H_X> + e^{i\Phi(t)}|V_{XX}V_X>,$$

with

$$\Phi(t) = (K_{V_{XX}} - K_{H_{XX}} + K_{V_X} - K_{H_X})*t_1 + (K_{V_X} - K_{H_X} + \frac{FSS}{\hbar})*t_2 +$$
$$+(K_{V_{XX}} - K_{H_{XX}})*(t_{prop}^{XX} - t_{start}^{XX}) + (K_{V_X} - K_{H_X})*(t_{prop}^X - t_{start}^X) +$$
$$+(\Phi_{V_{XX}}^0 - \Phi_{H_{XX}}^0 + \Phi_{V_X}^0 - \Phi_{H_X}^0)$$

From this, conditions *so that the phase term no longer depends on random variables $t_1$ and $t_2$* can be straightforwardly determined:

$$K_{V_X} - K_{H_X} = -\frac{FSS}{\hbar} \quad (5)$$

$$(K_{V_{XX}} - K_{H_{XX}}) = -(K_{V_X} - K_{H_X}) \quad (6)$$

If these conditions are met, the final 2 photon state resulting from the quantum dot emission after the compensation system will always be the same for each biexciton-exciton photon pair, as all of the other terms depend only on experimental parameters, such as the time of flight, that are assumed to remain constant over time: this implies that this residual constant phase term does not infringe on the resulting degree of entanglement, but merely changes the pure state all the photons pair will be entangled to, and can be easily compensated for using an additional static (i.e. non time-dependent) quantum gate.

Figure 1 provides a graphical and intuitive representation of the workings of the scheme we are proposing and of the conditions we have derived, in an ideal case where the other phase terms are simply assumed to be zero and the ramps are perfectly synchronized (these hypothesis are not necessary for the scheme to work, but merely simplify our analysis): condition (6) implies that the phase differences imparted between the two $|H_{XX}H_X>$ and $|V_{XX}V_X>$ by the operation on the biexciton line and that on the exciton line are equal but opposite. Provided the differential phase ramps are synchronized, the differential phase introduced on the biexciton photon emitted at $t_1$ will be perfectly compensated when the exciton photon, emitted at $t_1 + t_2$, crosses the compensation system. Likewise, condition (5) will ensure that the phase introduced by the evolution of the state due to the FSS is compensated as well when the exciton photon crosses the compensation system we have described. This removes the randomness of the final state.

Notably, one important advantage of this differential compensation approach is its insensitivity to any jitter present in the system, including those related to the excitation scheme employed, and to the starting point of the ramps, which as discussed before would only determine the final entangled state, but not the degree of entanglement (concurrence) of the source. Moreover, since all photon pairs will be transformed to the target state and none discarded, the technique we are proposing is in principle a lossless one: a major improvement with respect to time-gating techniques.

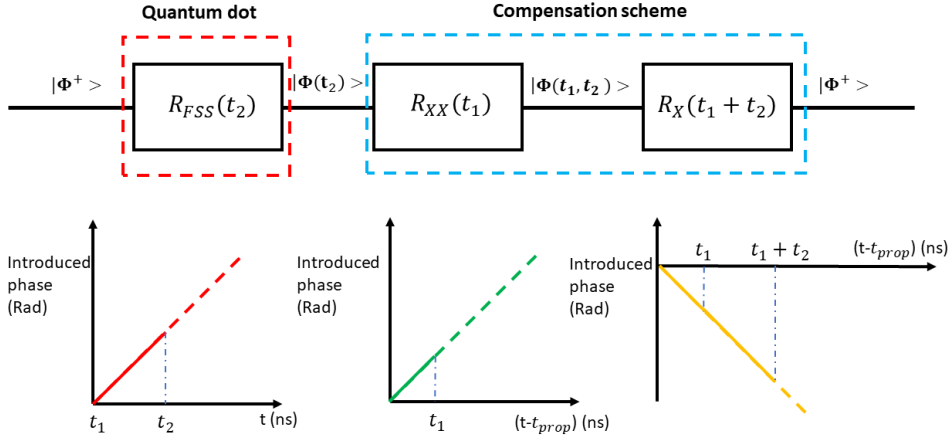

Figure 1 Circuit model representation of the overall temporally-phase evolution in the proposed compensation scheme: spin precession inside the quantum dot causes the build-up of a phase dependent on random time $t_2$. After emission, the effect of the scheme on the biexitonic part of the state adds another phase jump dependent on $t_1$ (emission time of the biexciton photon) when the XX photon crosses the system at time $t_1 - t_{prop}$. Both phase terms are compensated by the operation of the excitonic part of the wavefunction, if the proper conditions are met, when the exciton photon is imparted an opposite phase at time $t_1 + t_2 - t_{prop}$. Note that we have assumed $t_{prop}^{XX} = t_{prop}^{X} = t_{start}^{XX} = t_{start}^{X}$ for simplicity

We can also observe that from a quantum circuit perspective, the scheme hitherto proposed is equivalent to a cascade of two temporally-dependent phase gates, one operating on the biexciton and one operating on the exciton photon: in this model, the degradation of the entanglement due to the FSS is equivalent to the action of a phase gate $R_{FSS}(t_2)$, which turns what would otherwise be a perfect Bell state into a random one, $|\Phi(t_2)\rangle$ If the conditions we have derived are fulfilled, the sequential action of the two phase gates $R_{XX}(t_1)$ and $R_X(t_1 + t_2)$ would compensate the effect of the first phase gate, and result in the recovery of the original Bell state $|\Phi^+\rangle$. It should be noted that the scheme (and subsequent proposed implementation) we are proposing are somewhat similar to the one suggested by Wang et al. in their paper[30]. Their formalism however adopted a frequency domain approach, focussed on the removal of the energy difference between the two excitonic photon, and neglected the effect of other, time-independent phase terms on the outcome state. We believe our approach to provide a more natural, elegant, physically complete and simple to grasp picture, which especially in view of its circuit model interpretation could be easily applied to photonic quantum circuit design. Obviously, however, if one considers only frequency translation[37], the two can be seen as rather complementary: in fact, our solution will restore the energy degeneracy as well, and the two formalisms can be easily reconciled in the framework of signal theory using the concept of instantaneous angular frequency[38].

If we define the energy scale according to figure 2, we will have that:

$$\omega_X^V = \frac{E_X}{\hbar} + \frac{FSS}{2\hbar}$$

$$\omega_X^H = \frac{E_X}{\hbar} - \frac{FSS}{2\hbar}$$

(7)

For an angle modulated wave s(t)=Acos($\omega_c t + \vartheta(t)$)=Acos($\Phi$(t)), phase modulation will introduce an instantaneous angular frequency defined as $\omega_i(t) = \frac{d\Phi(t)}{dt}$, so that in our case, for a signal affected by the ramp, we will have:

$$\omega_X^V(t) = \omega_X^V + K_V$$

$$\omega_X^H(t) = \omega_X^H + K_H$$

(8)

And substituting equation 5 we will have that $\omega_X^V(t) - \omega_X^H(t) = 0$, which implies that the resulting photons will also be degenerate, and that the two approaches are consistent with each other. Indeed, the resulting state will be entangled in both polarization and energy, but we stress that, while the two effects (of phase correction and frequency translation) cannot be decoupled, what is needed to restore polarization entanglement is exclusively the phase correction, and the frequency translation should be regarded in principle as a side effect. Our discussion in the final part of our manuscript (see Fig 4) will help clarify that a simple frequency correction (ref 30) will not in general deliver a state whose fidelity to the ideal Bell state is unitary.

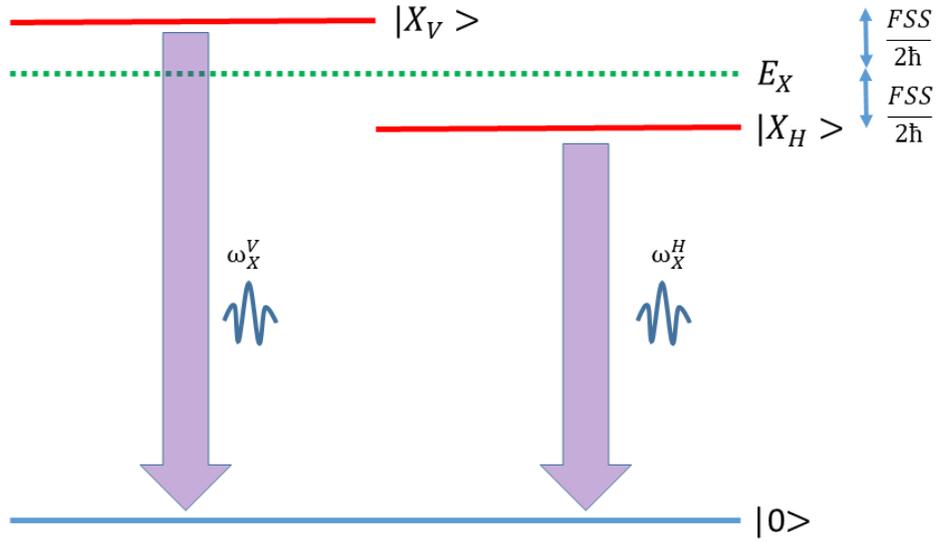

Figure 2 Graphical representation of the excitonic levels and fine-structure splitting. The zero in the energy scale is assumed to be the ground state

## Proposed implementation and feasibility analysis

In deriving the conditions that allow to restore the entanglement of the source, we have not made any assumption on the nature of the system that is supposed to implement them. It is quite obvious however that a natural realization of such a scheme could be based on the Pockels effect, as was also proposed in ref. [30]: a birefringent crystal would introduce a phase difference between polarizations aligned along the two optical axes, and the application of a transversal external electric field would allow to tune such a phase difference and change it in time by modifying the material's index ellipsoid: electro-optical phase modulators are indeed devices designed for this specific purpose, and that could be employed to implement our proposal.

For instance, in the most widely used material, $LiNbO_3$, the relationship between the applied field and the variation of the refractive index is described by the tensor[39]:

$$r_{i,k} = \begin{bmatrix} 0 & -r_{22} & r_{13} \\ 0 & r_{22} & r_{13} \\ 0 & 0 & r_{33} \\ 0 & r_{51} & 0 \\ r_{51} & 0 & 0 \\ -r_{22} & 0 & 0 \end{bmatrix} \qquad (9)$$

So that variations of the refractive index with the application of an external electric field can be written, in first approximation, as:

$$\Delta n_i = -\frac{n_i^3}{2}\sum_{k=1}^{3} r_{ik} E_k \tag{10}$$

Application of an external field in the z direction will affect the ordinary and extraordinary refractive indexes in a different way, owning to the different terms involved in the tensor product, and thus allow for an externally controlled (and potentially temporally-varying) phase gate.

Once the FSS, which is a property of the dot and can be measured experimentally is known, the relationship between the slope of the voltage ramp and the FSS can be easily determined in order to fulfill condition (5), while condition (6) can in turn be easily fulfilled by employing two phase modulators operated with opposite voltage ramps, one affecting the biexciton photon while the other affects the exciton, as shown in figure 3.

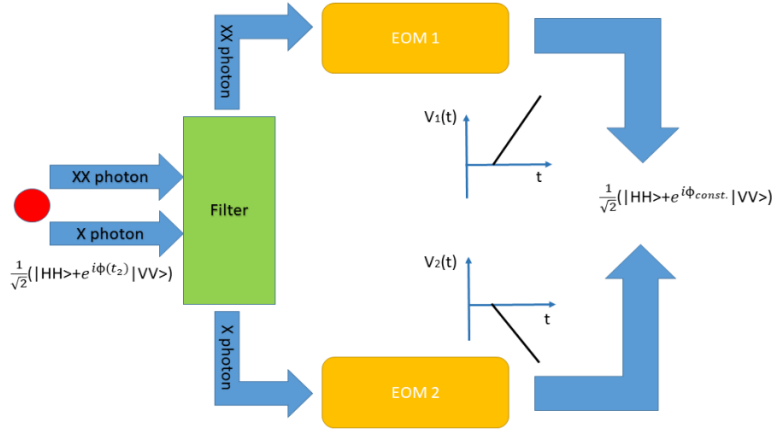

Figure 3 proposed implementation of the compensation scheme using two electo-optic modulators operated in a parallel configuration, as also envisioned by Wang et al.

A travelling wave, velocity matched phase modulator is conveniently characterized by its half-wave voltage $V_\pi$ that is, the value of the applied voltage that induces a phase shift of π on a wave travelling through it. As the electro-optic coefficients are different for the two axes, we can denote with $V_\pi^V$ and $V_\pi^H$ the values of $V_\pi$ for a V- and H-polarized wave respectively. The induced phase shift of a photon wave packet entering the modulator at time t for either polarization can then be easily written as $\Phi^{V,H}(t) = \frac{V(t)}{V_\pi^{V,H}} * \pi$

If the voltage is varying linearly in time, we can calculate the slope of induced phase by trivially taking the time derivative of this expression, and easily rewrite equation 5 as:

$$\frac{dV}{dt}\left(\frac{1}{V_\pi^V} - \frac{1}{V_\pi^H}\right) * \pi = -\frac{FSS}{\hbar}$$ to determine the slope of the voltage ramp.

If for instance we consider again a z-cut LiNbO3 integrated EOMs, and assume a $V_\pi$ of 3 Volts for the TM mode and a 3 times larger a $V_\pi$ for the TE modes, we will get a slope of the voltage ramp $\frac{dV}{dt}$ of around 2 V/(ns*µeV), which appears to be within the capabilities of modern high frequency EOMs if only a FSS of a few µeV has to be corrected, as it customarily found in dots grown on high symmetry substrates[13,14].

This implies that the proposed scheme can indeed be implemented with currently available technology, while advances in electro-optic material science[40] and integration[41] can pave the way for the development of compact quantum photonic circuits that perform a similar task and push the performance much further.

### Sensitivity to imperfections of the compensation scheme

In order to estimate the robustness of our approach to imperfections of the setup, a simple Monte Carlo simulation was performed, generating density matrixes for different values of the mismatches

$$\Delta\omega_1 = K_{V_X} - K_{H_X} - \frac{FSS}{\hbar}$$

$$\Delta\omega_2 = K_{V_{XX}} - K_{H_{XX}} + K_{V_X} - K_{H_X}$$

(11)

This was done by first of all assuming that $t_{prop}^{XX} = t_{start}^{XX}$ and $t_{prop}^{X} = t_{start}^{X}$, that is, a perfect syncing of the start of the ramp with the laser pulse, and by ignoring every other constant phase term. By creating random values for variables $t_1$ and $t_2$, a large number of density matrixes was generated and averaged until a suitable convergence was reached. In all of these simulations we have assumed a lifetime of the exciton state X of 1 ns, and half of that for the biexciton state, as it is often the case for III-V semiconductor quantum dots, and the FSS was chosen to be 3 µeV (corresponding to a spin precession of around 4.6 Rad/ns).

Figure 4 shows the behaviour of the concurrence of the resulting density matrixes ad their fidelity to the Bell state $|\Phi^+\rangle = \frac{1}{\sqrt{2}}(|HH\rangle + |VV\rangle) = \frac{1}{\sqrt{2}}(|00\rangle + |11\rangle)$ as a function of the errors $\Delta\omega_1$ and $\Delta\omega_2$: it can be observed that $\Delta\omega_2$ has a less severe impact due to the term being weighted by the random variable $t_1$, which is related to the shorter lifetime of the biexciton. For comparison, without the compensation scheme applied the same quantum dot would exhibit a fidelity and concurrence of 0.52 and 0.21 respectively. It can be also noticed that while $t_1$, the time that passes between the excitation of the quantum dot and the emission of the biexciton photon, plays no role in the level of the entanglement without our compensation scheme, it becomes important when the latter is applied: this is due to the differential phase gates being dependent on $t_1$ and $t_1 + t_2$.

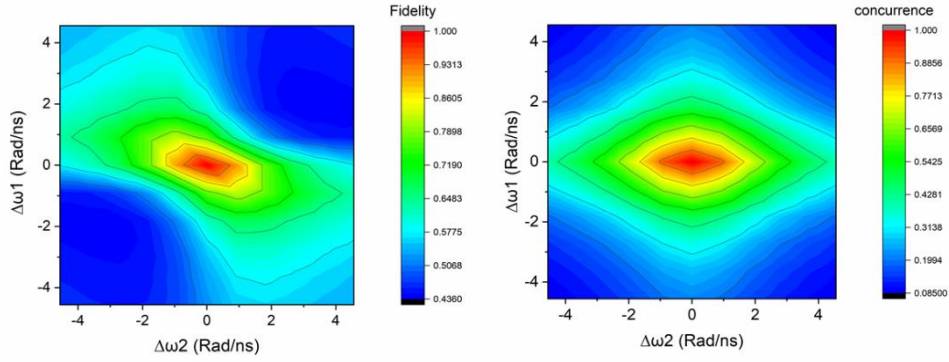

Figure 4 Fidelity (left) and Concurrence (right) of the state obtained after the compensation scheme for a dot with an FSS of 3 µeV, as a function of the errors $\Delta\omega_1$ and $\Delta\omega_2$. All of the other time-independent phase terms are assumed to be 1 for simplicity.

In order to also understand the impact of the constant phase term, a new series of simulations was performed. In this case, conditions to perfectly restore the entanglement of the state have been assumed to be fulfilled ($\Delta\omega_1 = \Delta\omega_2 = 0$), and concurrence and fidelity to the states $|\Phi^+>$ and $|\Phi^->$ were calculated as a function of the time delay δt= $\left(t_{prop}^{XX} - t_{start}^{XX}\right) - \left(t_{prop}^{X} - t_{start}^{X}\right)$, again for a quantum dot having a 3 µeV FSS.

As can be seen from figure 5, this has no effect on the level of concurrence, and thus on the degree of entanglement, but it can be exploited to finely tune the final state which is generated by the application of the compensation scheme.

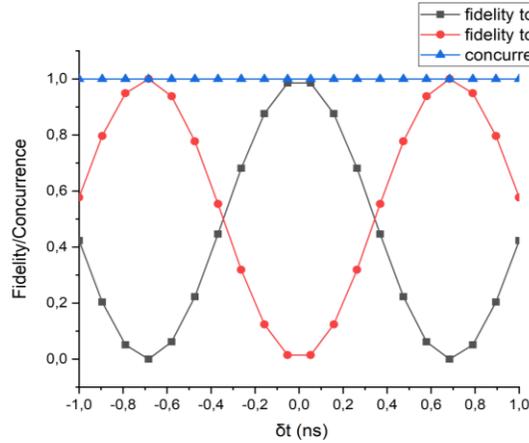

Figure 5 Fidelity to the Bell states and concurrence of the density matrix obtained after the compensation scheme as a function of temporal detuning δt in case of a perfect compensation of the FSS

In conclusion, we proposed a conceptually simple and intuitive way to compensate the effects of FSS on entanglement quality.

## Conclusions

The scheme we have discussed for restoring the entanglement of photons emitted from quantum dots using the biexciton-exciton recombination cascade appears to be quite robust, flexible and powerful, and it appears to be possible to implement it with currently available technology. As the implementation we have suggested is based on the repurposing of devices already widely employed in the (classical) field of information technology and telecommunications, it represents an elegant solution that could significantly lower infrastructural costs associated with a "quantum internet". This is especially true since in principle no photon losses are introduced by our approach, which would result in a higher bitrate Moreover, by introducing the paradigm shift of a-posteriori compensation of the effect of the FSS, constraints on the sources that could be employed for such a purpose are significantly relaxed, and fabrication of devices integrating semiconductor quantum dots as entangled photon sources for quantum information processing are simplified.

## Methods

Monte Carlo simulations were performed using a custom-written Python code. Large numbers of density matrixes were generated and averaged until a suitable convergence was reached, defined as a less than $10^{-6}$ relative change in the matrix elements from the previous iteration.

Fidelity of the converged density matrixes to any Bell state was calculated as:

$$F = \left( Tr\left( \sqrt{\sqrt{\rho_B} * \rho * \sqrt{\rho_B}} \right) \right)^2$$

Where we have used $\rho_B$ and $\rho$ to indicate the density matrix of the target Bell state and the one resulting from our simulation, respectively.

Concurrence was calculated using the standard formula:

$$C(\rho) = \max\left(0, \lambda_1 - \lambda_2 - \lambda_3 - \lambda_4\right)$$

With $\lambda_n$ being the squares roots of the eigenvalues, in decreasing order, of the operator $R = \rho \Sigma \rho^T \Sigma$, and $\Sigma = \begin{pmatrix} 0 & 0 & 0 & -1 \\ 0 & 0 & 1 & 0 \\ 0 & 1 & 0 & 0 \\ -1 & 0 & 0 & 0 \end{pmatrix}$

## Author contribution

S.V. conceived the idea and developed the code used to simulate the density matrixes with assistance from G.J. E.P supervised the work and assisted in writing the manuscript. All authors reviewed the manuscript.

## Acknowledgments

This research was supported by Science Foundation Ireland under Grant Nos. 15/IA/2864, SFI-18/SIRG/5526, and 12/RC/2276_P2 .

**Additional Information**

The authors declare no competing interests.

**Data availability**

Our simulations have been performed using a custom-written code in the 3.7 version of the Python language. The code uses the *Scipy* and *Numpy* Python libraries, and was run via the *Anaconda* Python distribution, with typical computation times of less than 2 hours on a 3.4 GHz quad-core Intel i3 machine with 16 Gbyte of RAM. All code is available free of charge upon request.